\documentclass[%
preprint,
showpacs,
 amsmath,amssymb,
 aps,
]{revtex4-1}

\newcommand{\gv}[1]{\ensuremath{\mbox{$ #1 $}}} 
\newcommand{\uv}[1]{\ensuremath{\mathbf{\hat{#1}}}} 
 
\newcommand{\grad}[1]{\gv{\nabla} #1} 
\newcommand{\curl}[1]{\gv{\nabla} \times #1} 
\let\baraccent=\= 
\renewcommand{\=}[1]{\stackrel{#1}{=}} 

\usepackage{amsfonts}
\usepackage{amsmath}
\usepackage{amssymb}
\usepackage{gensymb}
\usepackage{commath}
\usepackage{color}
\usepackage{makeidx}
\usepackage{graphicx}
\usepackage{adjustbox}
\usepackage{siunitx}
\usepackage{epstopdf}
\usepackage{caption}
\usepackage{setspace}
\usepackage{subcaption}
\usepackage{graphicx}
\usepackage{dcolumn}
\usepackage{bm}
\usepackage{easybmat}
\usepackage{hyperref}

\usepackage{xcolor}
\hypersetup{
    colorlinks,
    linkcolor={blue},
    citecolor={blue},
    urlcolor={blue}
}

\usepackage{natbib}
\begin{document}


\title{Effective dielectric tensor of deformed-helix ferroelectric liquid crystals with subwavelength pitch and large tilt angle}

\author{Leonardo Silvestri}
 \email{l.silvestri@unsw.edu.au}
\affiliation{%
 School of Electrical Engineering and Telecommunications, UNSW Sydney, Australia
}%
\author{Hrishikesh Srinivas}
\thanks{Now at the Department of Electrical Engineering, Stanford University, USA}
\affiliation{%
 School of Electrical Engineering and Telecommunications, UNSW Sydney, Australia
}%
\author{Fran\c{c}ois Ladouceur}%
\affiliation{%
 School of Electrical Engineering and Telecommunications, UNSW Sydney, Australia
}%

\date{\today}

\begin{abstract}
Short pitch deformed helix ferroelectric liquid crystals have numerous applications as active materials in displays, optical telemetry and biomedical devices. In this paper, we derive convenient analytical formulas to calculate the effective dielectric tensor of these materials beyond the space average approximation. By comparison with exact numerical calculations, we show that our formulas are remarkably accurate in predicting optical properties in virtually all practical situations, including the important case of large tilt angles, where the space average approximation breaks down. We also present a comparison between the two complementary approaches of expanding the mesoscopic dielectric tensor vs. the mesoscopic transfer matrix, by deriving an expression for the effective transfer matrix as an infinite expansion and explicitly calculating the corresponding effective dielectric tensor for the first time. Our results demonstrate that both methods give accurate predictions when two-photon scattering terms are taken into account.
\end{abstract}

\pacs{61.30.Gd, 61.30.Hn,77.84.Nh,42.79.Kr}
                     
\maketitle

\section{\label{sec:level1}Introduction}
Liquid crystals (LCs) are extensively used as active elements in displays \cite{Kim:2009aa} and optical communication devices \cite{Hirabayashi_1993,DeBougrenet_2004}, thanks to the fact that their optical properties can be controlled by applying external electric fields. More recently, LCs have enjoyed a renaissance and found new applications in biology \cite{Abdulhalim_bio_2011,Woltman:2007aa} and nano-technology \cite{LAGERWALL20121387}. In particular, deformed-helix ferroelectric liquid crystals (DHFLCs) are a class of materials \cite{Beresnev_LC_1989}, that are being investigated for application not only to displays \cite{Hegde_LC_2008}, but also to optical sensing networks \cite{Firth2017Loop,firth2016flow,Firth2017HV}, Q-switched \cite{Wieschendorf20171692} and mode-locked \cite{Lei:18} lasers, and biomedical devices \cite{Al-Abed2018Optrode,Ladouceur:18}. Homogeneously aligned DHFLCs have been shown to exhibit a remarkably linear and fast response to very small electric fields \cite{brodzeli2013sensors}, characteristics that make them suitable to detection of sub-mV voltages, such as those in biological tissues \cite{Al-Abed2018Optrode}. 
Given the number of practical applications, it is important to develop accurate models to simulate the electro-optic response of these materials and optimize the performance of the various devices. Fully numerical simulations are now possible, {\em e.g.} using finite element analysis software, where the electro-optical properties of a LC cell can be calculated from the microscopic parameters of the material. However, it is often desirable to have a simpler analytical model that can clearly explain the relation between parameters, even if at the cost of approximating the problem. 
We are particularly interested in describing the reflection and transmission of cells containing homogeneously aligned DHFLCs liquid crystals driven by an external electric field, as this is a technologically relevant configuration. It has been shown that in the limit of short pitch, {\em i.e.} when the pitch is much smaller than the wavelength of the incident light, an effective medium description is appropriate \cite{Oldano_PRB_96,Galatola_97}, because the only meaningful physical quantities are the averages of the fields over a scale that is larger than the pitch and smaller than the wavelength \cite{Landau_Lifshitz}. In particular, the effective medium approximation works well for the homogeneous alignment considered in this paper, even when spatial dispersion, and therefore optical activity, are considered \cite{Galatola_97}. Several approaches have been put forward to determine a suitable effective dielectric tensor and we can broadly subdivide them into two categories: methods relying on a perturbative expansion of the dielectric tensor, such as the Bloch wave method \cite{Galatola_97}, and methods relying on a perturbation of the transfer matrix, such as the iteration procedure by Oldano et al. \cite{Oldano_PRB_96, Becchi_JOA_99} and the polarization grating approach \cite{Kiselev2011, Pozhidaev_2013,Kiselev_2014}. 
Galatola \cite{Galatola_97} has presented a general expression for the effective dielectric tensor based on a Bloch-wave decomposition and applied it to the case of cholesteric LCs. Oldano et al. \cite{Oldano_PRB_96,Becchi_JOA_99} have reported a method to iteratively calculate an effective transfer matrix in the case of samples placed between parallel planes orthogonal to the direction of periodicity of the dielectric medium. They have also shown that, for this particular orientation, optical activity cannot be described by an effective dielectric tensor. However, their method does not apply to the homogeneously aligned DHFLCs considered here. Other groups have derived explicit expressions for the space averaged effective transfer matrix and reported the corresponding effective dielectric tensor, including explicitly the effect of an external electric field \cite{Kiselev2011,brodzeli2013reflective}, but we will show that this approximation is too crude to describe DHFLCs with large tilt angles.

In this paper we present a comparison between the two complementary approaches of expanding the mesoscopic dielectric tensor vs. the mesoscopic transfer matrix, filling the gaps in the literature and using the same conceptual framework for both methods.
In particular, we derive the following results: (i) an explicit expansion for the effective transfer matrix, (ii) two explicit expansions for the effective dielectric tensor, obtained with the two above mentioned approaches, (iii) simple analytical formulas for the effective dielectric tensor of short pitch, electrically driven, homogeneously aligned DHFLCs, which are extremely accurate within the whole range of practical material parameters. We then compare our analytical expressions to the exact numerical treatment. We anticipate that the two methods give similar accuracy and that just a few terms of the infinite expansions are generally sufficient to describe even LCs with large tilt angles in the whole range of realistic parameters.
We also find that, while the transfer matrix approach method is more similar, in spirit, to the exact numerical approach we adopted, the Bloch wave method yields more convenient formulas, which will certainly be useful to the many researchers working in the area of DHFLCs. We point out that optical activity is not discussed, as we focus on the zeroth order of the short pitch expansion, where spatial dispersion is neglected.
The paper is organized as follows. In the next section we describe in detail the transfer matrix approach for a generic dielectric grating consisting of a slab with its dielectric tensor periodic along a direction parallel to the slab surface. In Section III, we present the analytical results for the same geometry, derived using the Bloch wave approach. In Section IV, the previous results are used to derive explicit formulas for the effective dielectric tensor of DHFLCs. Finally in Section V, we compare the analytical results with an exact numerical approach. The last section is devoted to the conclusions.

\section{\label{sec:level1}Transfer matrix approach}
\subsection{\label{sec:level2}Problem definition}

\begin{figure}[h]
\includegraphics[width=15cm]{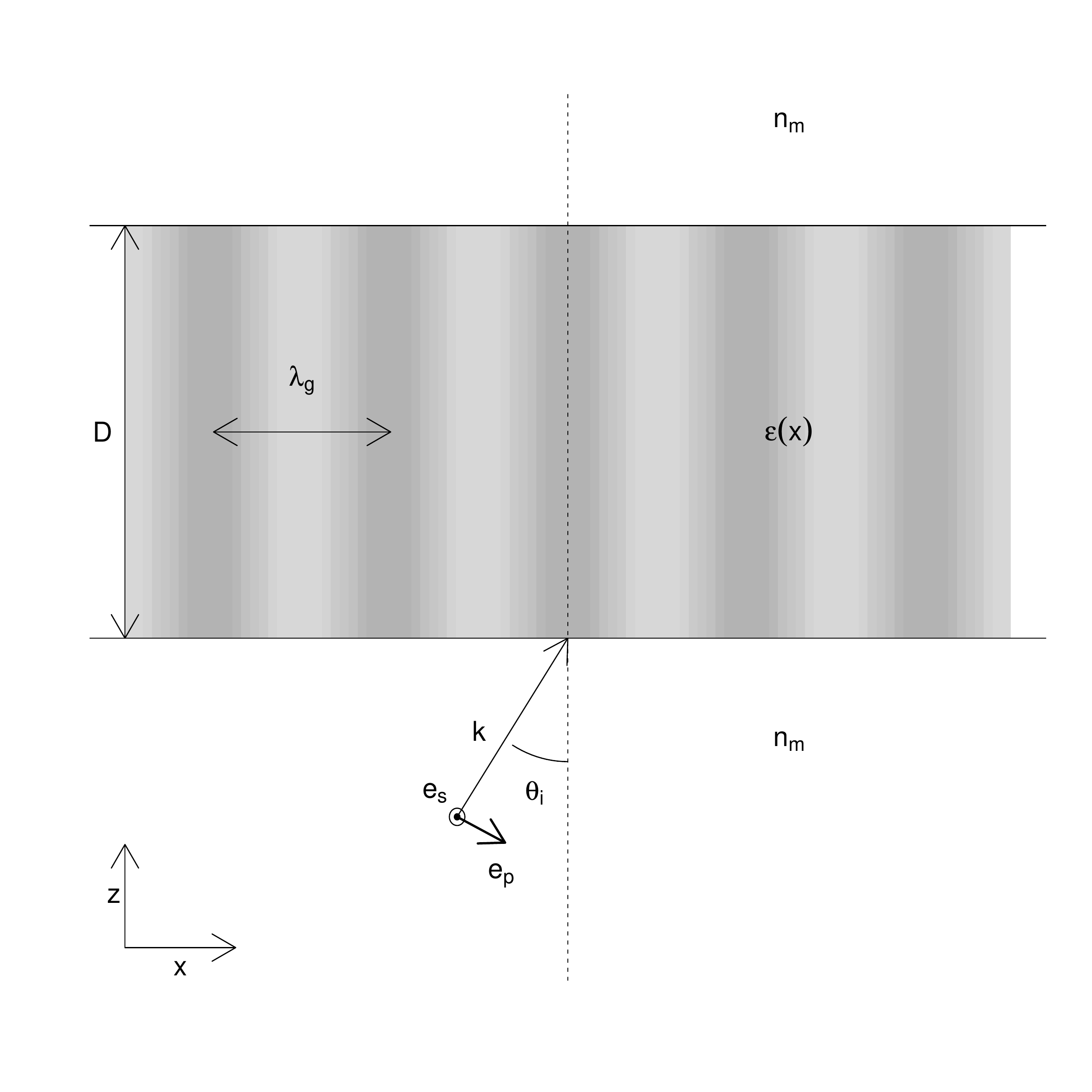}
\caption{Geometry of the problem. A planar dielectric grating of thickness $D$ is placed in an isotropic ambient medium with refractive index $n_{\text{m}}$. The dielectric tensor of the grating, $\boldsymbol{\epsilon}(x)$, varies periodically in a direction parallel to the slab's surface with pitch $\lambda_{\text{g}}$. This geometry describes deformed helix ferroelectric liquid crystals with homogeneous alignment.}
\label{Fig_geometry}
\end{figure}

We consider the problem of a dielectric grating in a slab geometry oriented as in Figure \ref{Fig_geometry}, where the dielectric constant is periodically varying along $\hat{x}$. 
We begin by considering harmonic electromagnetic waves of the form $\mathbf{E}(\mathbf{r},t) = \mathbf{E}(\uv{k})\*\exp(i\*[\mathbf{k} \cdot \mathbf{r}-\omega\*t])$, which satisfy Maxwell's equations
\begin{subequations}
\begin{align}
\curl{\mathbf{E}} &= i\*k_{0}\*\boldsymbol{\mu}\mathbf{H}\\
\curl{\mathbf{H}} &= -i\*k_{0}\*\boldsymbol{\epsilon}\mathbf{E},
\end{align}
\end{subequations}
where $\{\mathbf{E},\mathbf{H}\}$ is the electric and magnetic field-strength vector pair, $k_{0} = 2\*\pi/\lambda_{0}$ is the wavenumber of light of wavelength $\lambda_{0}$ in free space, $\boldsymbol{\mu}$ is the magnetic tensor and $\boldsymbol{\epsilon}$ the dielectric tensor, both of rank 2. The wave vector in an isotropic ambient medium (denoted by subscript m) can be written in terms of a unit wave vector as $\mathbf{k} = k_{\text{m}}\*\uv{k}$, where $k_{\text{m}} = n_{\text{m}}\*k_{0}$ with $n_{\text{m}} = \sqrt{\mu_{\text{m}}\*\epsilon_{\text{m}}}$. For uniform plane waves we neglect time-dependence, so that $\{\mathbf{E},\mathbf{H}\}=\{\mathbf{E}(\uv{k}),\mathbf{H}(\uv{k})\} \*\exp(i\*\mathbf{k} \cdot \mathbf{r})$.
A wave thus defined propagates in an ambient medium for $z < 0$ and impinges on a dielectric slab of thickness $D$ at the interface $z = 0$ at an angle of incidence $\theta_{\text{i}}$ to the normal in the plane of incidence, which is spanned by $\uv{k} = \left( \sin{\theta_{\text{i}}}\*\cos{\varphi_{\text{i}}},\ \sin{\theta_{\text{i}}}\*\sin{\varphi_{\text{i}}},\ \cos{\theta_{\text{i}}} \right)$ and $\mathbf{e}_{p}(\uv{k}) = \left( \cos{\theta_{\text{i}}}\*\cos{\varphi_{\text{i}}},\ \cos{\theta_{\text{i}}}\*\sin{\varphi_{\text{i}}},\ -\sin{\theta_{\text{i}}} \right)$. The azimuthal angle $\varphi_{\text{i}}$ is the angle formed between the plane of incidence and the $xz$ plane (the grating plane). Hence, with $\mathbf{e}_{s}(\uv{k}) = \left( -\sin{\varphi_{\text{i}}},\ \cos{\varphi_{\text{i}}},\ 0 \right)$, 
\begin{subequations}
\begin{align}
\mathbf{E}(\uv{k}) &= E_{\parallel}\*\mathbf{e}_{p}(\uv{k})+E_{\perp}\*\mathbf{e}_{s}(\uv{k}) \\
\mu_{\text{m}}\*\mathbf{H}(\uv{k}) = k_{0}^{-1}\*\mathbf{k}\times \mathbf{E} &= n_{\text{m}}\*\left[ E_{\parallel}\*\mathbf{e}_{s}(\uv{k})-E_{\bot}\mathbf{e}_{p}(\uv{k}) \right].
\end{align}
\label{fields}
\end{subequations}
In the case considered here, the magnetic tensor is assumed to be isotropic, while the dielectric tensor is assumed to be symmetric and periodic along $x$ with period $\lambda_{\text{g}}$. In formulas
\begin{subequations}
\begin{align}
\mu_{ij} &= \mu\*\delta_{ij}\\
\epsilon_{ij} &= \epsilon_{ji}\\
\epsilon_{ij}(x+\lambda_{\text{g}})&=\epsilon_{ij}(x).
\end{align}
\end{subequations}
The quantity $\lambda_{\text{g}}$ is called the grating pitch and the corresponding grating wavenumber is defined as $k_{\text{g}}=2\pi / \lambda_{\text{g}}$.
We assume further the formalism of \citet{Kiselev2011} for Maxwell's equations in terms of the lateral components of the electromagnetic field $\mathbf{E}_{\text{p}}$, $\mathbf{H}_{\text{p}}$, defined by
\begin{align}
\mathbf{E} = E_{z}\*\uv{z}+\mathbf{E}_{\text{p}}, \quad \mathbf{H} = H_{z}\*\uv{z}+\uv{z} \times \mathbf{H}_{\text{p}},
\label{lat}
\end{align}
or equivalently, $\mathbf{E}_{\text{p}} = \begin{pmatrix} E_{x}, E_{y} \end{pmatrix}^{T}$ and $\mathbf{H}_{\text{p}} = \begin{pmatrix} H_{y}, -H_{x} \end{pmatrix}^{T}$, where superscript $T$ denotes the matrix transpose. The wave vector may be decomposed in similar fashion as $\mathbf{k} = k_{z}\*\uv{z}+\mathbf{k}_{\text{p}}$, where $\mathbf{k}_{\text{p}} = \sqrt{k_{\text{m}}^{2}-k_{z}^{2}}\*\left(\cos{\varphi_{\text{i}}} \ \*\uv{x}+\sin{\varphi_{\text{i}}}\* \ \uv{y}\right)$ is the lateral wave-vector component of the incident wave, which remains constant for all propagation by the appropriate boundary conditions for continuity of $\mathbf{E}_{\text{p}}$, $\mathbf{H}_{\text{p}}$ \cite{Kiselev2008}. 
The resulting system of differential equations may be written in simplified form, with $\tau = k_{0}\*z$, as
\begin{align}
-i\partial_{\tau} \mathbf{F} = \hat{\mathcal{M}}\*\mathbf{F} = \begin{pmatrix} \hat{\mathcal{M}}^{(11)} & \hat{\mathcal{M}}^{(12)} \\ \hat{\mathcal{M}}^{(21)} & \hat{\mathcal{M}}^{(22)}\end{pmatrix} \* \begin{pmatrix} \mathbf{E}_{\text{p}}\\ \mathbf{H}_{\text{p}}\end{pmatrix}.
\label{system}
\end{align}
$\hat{\mathcal{M}}_{\alpha \beta}^{(ij)}$ are the differential operators
\begin{subequations}
\begin{align}
\hat{\mathcal{M}}_{\alpha \beta}^{(11)} &= -\grad_{\alpha} \epsilon_{zz}^{-1}\*\epsilon_{z\beta} \\
\hat{\mathcal{M}}_{\alpha \beta}^{(12)} &= \mu\*\delta_{\alpha \beta}-\grad_{\alpha} \epsilon_{zz}^{-1}\*\grad_{\beta} \\
\hat{\mathcal{M}}_{\alpha \beta}^{(21)} &= \epsilon_{\alpha \beta} - \epsilon_{\alpha z}\*\epsilon_{zz}^{-1}\*\epsilon_{z \beta} - \mu^{-1}\*\grad_{\alpha}^{\bot}\*\grad_{\beta}^{\bot} \\
\hat{\mathcal{M}}_{\alpha \beta}^{(22)} &= -\epsilon_{\alpha z}\*\epsilon_{zz}^{-1}\grad_{\beta}
\end{align}
\end{subequations}
with $\alpha, \beta \in \{x,y\}$, and ($x,y$ interchangeably $1,2$)
$$\grad_{\alpha} = \uv{\boldsymbol{\alpha}}\*\frac{1}{i\*k_{0}}\*\frac{\partial}{\partial \alpha},
\qquad 
\grad_{\alpha}^{\bot} = -\uv{\boldsymbol{\alpha}}\*\frac{\varepsilon_{\alpha \beta}}{i\*k_{0}}\*\frac{\partial}{\partial \beta}$$
where $\varepsilon_{\alpha \beta}$ is the two-dimensional Levi-Civita symbol, and the Einstein summation convention applies to $\beta$ in the second expression.  

\subsection{\label{sec:level2}Floquet Diffraction Harmonics}

The in-plane periodicity of the dielectric tensor allows its expansion as a Fourier series
\begin{align}
\epsilon_{ij} =\sum_{n=-\infty}^{\infty} \epsilon_{ij,n}\exp(i\*n\*k_{\text{g}} x).
\end{align}
The $\hat{\mathcal{M}}_{\alpha \beta}^{(ij)}$ are thus periodic, so that \eqref{system} is a differential problem amenable to Floquet theory \cite{Floquet1883}, allowing a Floquet harmonics representation for the solution as
\begin{align}
\begin{split}
\mathbf{F}(\mathbf{r}) &= \mathbf{F}(\mathbf{r}_{\text{p}}, \tau) \\
&= \sum_{n=-\infty}^{\infty} \mathbf{F}_{n}(\tau) \*\exp(i\*\mathbf{k}_{n} \cdot \mathbf{r}_{\text{p}}),
\label{soln}
\end{split}
\end{align}
where $\mathbf{r}_{\text{p}} = (x, y, 0)$, and $\mathbf{k}_{n} = \mathbf{k}_{\text{p}}+n\*\mathbf{k}_{\text{g}} = k_{0}\*\mathbf{q}_{n}$, with the dimensionless wave index vector $\mathbf{q}_{n} = (q_{x}^{(n)},q_{y}^{(n)},0)$, and $q_{n} = |\mathbf{q}_{n}|$. From the definitions introduced, we note
\begin{subequations}
\begin{align}
q_{x}^{(n)} = q_{n}\*\cos{\phi_{n}} &= n_{\text{m}}\*\sin{\theta_{\text{i}}}\*\cos{\varphi_{\text{i}}}+n\*\frac{k_{\text{g}}}{k_{0}} \\
q_{y}^{(n)} = q_{n}\*\sin{\phi_{n}} &= n_{\text{m}}\*\sin{\theta_{\text{i}}}\*\sin{\varphi_{\text{i}}},
\end{align}
\label{q}
\end{subequations}
so that the rotation angle $\phi_{n}$ diminishes with increasing order $n$. We will also make use of the vector $\mathbf{p}_{n} = \uv{z} \times \mathbf{q}_{n}$.
Substituting \eqref{soln} in \eqref{system} and taking the inner product on both sides with the Fourier kernel function, we obtain the system of equations for Floquet harmonics $\mathbf{F}_{n}(\tau)$,
\begin{align}
-i\partial_{\tau} \mathbf{F}_{n}(\tau) = \sum_{m=-\infty}^{\infty} \mathbf{M}_{nm}(\tau)\*\mathbf{F}_{m}(\tau)
\label{newsys}
\end{align}
where 
\begin{equation}
\mathbf{M}_{nm}=\left(\begin{matrix} \mathbf{M}_{nm}^{(11)} & \mathbf{M}_{nm}^{(12)} \\ \mathbf{M}_{nm}^{(21)} & \mathbf{M}_{nm}^{(22)} \end{matrix}\right).
\end{equation}
$[\mathbf{M}_{nm}^{(ij)}]_{\alpha \beta}$ are the $n$th order Fourier coefficients of the 2$\times$2 matrices $\hat{\mathcal{M}}_{\alpha \beta}^{(ij)}\*\exp(i\*\mathbf{k}_{m} \cdot \mathbf{r})$, given explicitly by 
\begin{subequations}
\begin{align}
[\mathbf{M}_{nm}^{(11)}]_{\alpha \beta} & =  -q_\alpha^{(n)} \beta_{z\beta}^{(n-m)}, \\
{[\mathbf{M}_{nm}^{(12)}]}_{\alpha \beta} & =  \mu \delta_{\alpha\beta}\delta_{nm}-q_\alpha^{(n)}\eta_{zz}^{(n-m)}q_\beta^{(m)} , \\
{[\mathbf{M}_{nm}^{(21)}]}_{\alpha \beta} & =  \xi_{\alpha\beta}^{(n-m)}-\mu^{-1}\delta_{nm}p_\alpha^{(n)}p_\beta^{(m)} , \\
{[\mathbf{M}_{nm}^{(22)}]}_{\alpha \beta} & =  -\beta_{\alpha z}^{(n-m)} q_\beta^{(m)},
\end{align}
\label{M_Fourier}
\end{subequations}
in terms of the $n$th order Fourier coefficients 
\begin{subequations}
\begin{align}
\xi_{\alpha \beta}^{(n)} & = \frac{1}{\lambda_{\text{g}}} \int_0^{\lambda_{\text{g}}} e^{-i n k_{\text{g}} x} \left[\epsilon_{\alpha \beta} - \epsilon_{\alpha z}\*\epsilon_{zz}^{-1}\*\epsilon_{z \beta}\right] {\rm d}x,\\
\eta_{zz}^{(n)} & = \frac{1}{\lambda_{\text{g}}} \int_0^{\lambda_{\text{g}}} e^{-i n k_{\text{g}} x} \left[ \epsilon_{zz}^{-1} \right] {\rm d}x, \\
\beta_{\alpha z}^{(n)} & = \frac{1}{\lambda_{\text{g}}} \int_0^{\lambda_{\text{g}}} e^{-i n k_{\text{g}} x} \left[ \epsilon_{zz}^{-1}\*\epsilon_{\alpha z} \right] {\rm d}x, \\
\beta_{z\alpha }^{(n)} & = \frac{1}{\lambda_{\text{g}}} \int_0^{\lambda_{\text{g}}} e^{-i n k_{\text{g}} x} \left[ \epsilon_{zz}^{-1}\*\epsilon_{z\alpha} \right] {\rm d}x.
\end{align}
\label{Fourier_coefficient}
\end{subequations}
It is straightforward to verify that, due to the dielectric tensor symmetry, we have the following relations for the transfer matrix terms
\begin{subequations}
\begin{align}
\beta_{\alpha z}^{(n)} & = \beta_{z \alpha}^{(n)}, \\
\beta_{\alpha z}^{(-n)} & = [\beta_{\alpha z}^{(n)}]^*, \\
\xi_{\alpha \beta}^{(n)} & =\xi_{\beta \alpha}^{(n)}, \\
\xi_{\alpha \beta}^{(-n)} & = [\xi_{\alpha\beta}^{(n)}]^*, \\ 
\eta_{zz}^{(-n)} & = [\eta_{zz}^{(n)}]^*.
\end{align}
\label{Fourier_coefficient_properties}
\end{subequations}
In the ambient medium, Floquet harmonics are decoupled and represent forward and backward-propagating eigenwaves with wave vector $z$-components $k_{z}^{\pm} = \pm k_{0}\* \sqrt{n_{\text{m}}^{2} - q_{n}^{2}}$, respectively. From the inequality
\begin{align}
q_{n} \geq n\*\frac{k_{\text{g}}}{k_{0}} = n\*\frac{\lambda_{\text{0}}}{\lambda_{\text{g}}},
\end{align}
it follows that in the limit of gratings with very short pitch, {\em i.e.} as $\lambda_{\text{g}}/\lambda_{0} \rightarrow 0$, all diffraction orders with $n \geq 1$ correspond to evanescent waves, more strongly attenuated with increasing order. This also applies to Floquet harmonics inside the slab and it creates the possibility of defining an effective tensor to describe the optical properties of the grating, valid in the limit $\lambda_{\text{g}}/\lambda_{0} \rightarrow 0$. However, it is important to recognize the fact that in the slab all Floquet harmonics are coupled by the presence of the grating. 
\subsection{\label{sec:level2}Effective transfer matrix and effective dielectric tensor} 
In this section we derive the analytical expression of an effective transfer matrix, $\mathbf{M}^{(\rm{eff})}$, that correctly describes the optical properties of the polarization grating in the limit $\lambda_{\text{g}}/\lambda_{0} \rightarrow 0$. We use a procedure analogous to time-independent perturbation theory in quantum mechanics, where the transfer matrix $\boldsymbol{\hat{\mathcal{M}}}$ plays the role of the Hamiltonian. We split $\boldsymbol{\hat{\mathcal{M}}}$ into a unperturbed part, $\boldsymbol{\hat{\mathcal{M}}}_0$, describing an average transfer matrix, and a perturbation, $\boldsymbol{\Delta\hat{\mathcal{M}}}(x)$, due to the grating and depending on the spatial coordinate $x$. Once expressed in terms of Floquet harmonics, we have 
\begin{equation}  
\boldsymbol{\mathcal{M}}=\boldsymbol{\mathcal{M}}_0+\boldsymbol{\Delta \mathcal{M}}
\end{equation}
\begin{align}
\boldsymbol{\mathcal{M}}_0 = \begin{pmatrix} 
\ddots & \vdots &  \vdots & \vdots   &  \\
\cdots & \mathbf{M}_{-1 -1} & \mathbf{0} & \mathbf{0} & \cdots \\ 
\cdots & \mathbf{0} & \mathbf{M}_{0 0} &  \mathbf{0} &  \cdots \\   
\cdots & \mathbf{0} & \mathbf{0} & \mathbf{M}_{1 1}  &  \cdots \\  
 & \vdots &  \vdots & \vdots   & \ddots \\
\end{pmatrix},
\label{M0}
\end{align}
\begin{align}
\boldsymbol{\Delta \mathcal{M}} = \begin{pmatrix} 
\ddots & \vdots &  \vdots & \vdots   &  \\
\cdots & \mathbf{0} & \mathbf{M}_{-1 0} & \mathbf{M}_{-1 1} & \cdots \\ 
\cdots & \mathbf{M}_{0 -1} & \mathbf{0} &  \mathbf{M}_{0 1} &  \cdots \\   
\cdots & \mathbf{M}_{1 -1} & \mathbf{M}_{1 0} & \mathbf{0}  &  \cdots \\   & \vdots &  \vdots & \vdots   & \ddots \\
\end{pmatrix},
\label{DeltaM}
\end{align}
Note that $\boldsymbol{\mathcal{M}}$, $\boldsymbol{\mathcal{M}}_0$ and $\boldsymbol{\Delta \mathcal{M}}$ are non-Hermitian and that the Floquet harmonics basis consists of 4-dimensional vectors, so that the elements of the above matrices are non-Hermitian, non-commuting 4x4 matrices. Keeping it in mind, we can still adopt a procedure similar to time-independent quantum mechanics perturbation theory. The ``eigenvalues" of the unperturbed transfer matrix $\boldsymbol{\mathcal{M}}_0$ are the 4$\times$4 transfer matrices $\mathbf{M}_{n n}$, describing the evolution of waves in the absence of any spatial modulation of the dielectric tensor. Our goal is to find the perturbed ``eigenvalue" of the 0$^{\rm th}$ order Floquet harmonic, corresponding to the evolution of the four propagating waves. After a tedious derivation, we obtain, up to the third order in the perturbation,
\begin{widetext}
\begin{equation}\label{M_eff_expansion}
\mathbf{M}^{(\rm{eff})}_{\lambda_{\text{g}}} = \mathbf{M}_{0 0}-\sum_{n \ne 0} \mathbf{M}_{0 n}\left[ \mathbf{M}_{n n} \right]^{-1} \mathbf{M}_{n 0}+\sum_{n \ne 0}\sum_{m \ne \{0,n\}} \mathbf{M}_{0 n}\left[ \mathbf{M}_{n n} \right]^{-1} \mathbf{M}_{n m} \left[ \mathbf{M}_{m m} \right]^{-1} \mathbf{M}_{m 0},
\end{equation}
\end{widetext}
which can be explicitly calculated from equations (\ref{M_Fourier}) and depends on the grating pitch $\lambda_{\text{g}}$ through $\mathbf{q}_n$. By taking the short pitch limit, we get the effective 4x4 transfer matrix 
\begin{equation}
\mathbf{M}^{(\rm{eff})}=\lim_{\lambda_{\text{g}} \rightarrow 0} \mathbf{M}^{(\rm{eff})}_{\lambda_{\text{g}}},
\end{equation}
which is finite and describes a homogeneous medium. The effective dielectric tensor is then defined by the relations
\begin{subequations}
\begin{align}
\label{M_eff_Fourier_a}
[\mathbf{M}^{(\rm{eff},11)}]_{\alpha \beta} & =  -q_\alpha^{(0)} \frac{\epsilon^{(\rm{eff})}_{z\beta}}{\epsilon^{(\rm{eff})}_{zz}}, \\ \label{M_eff_Fourier_b}
{[\mathbf{M}^{(\rm{eff},12)}]}_{\alpha \beta} & =  \mu \delta_{\alpha\beta}-\frac{q_\alpha^{(0)}q_\beta^{(0)}}{\epsilon^{(\rm{eff})}_{zz}} , \\ \label{M_eff_Fourier_c}
{[\mathbf{M}^{(\rm{eff},21)}]}_{\alpha \beta} & =  \epsilon_{\alpha \beta}^{(\rm{eff})} -\frac{ \epsilon_{\alpha z}^{(\rm{eff})}\epsilon_{z \beta}^{(\rm{eff})}}{\epsilon_{zz}^{(\rm{eff})}}-\mu^{-1}p_\alpha^{(0)}p_\beta^{(0)} , \\
{[\mathbf{M}^{(\rm{eff},22)}]}_{\alpha \beta} & =  -\frac{\epsilon^{(\rm{eff})}_{\alpha z}}{\epsilon^{(\rm{eff})}_{zz}} q_\beta^{(0)},
\end{align}
\label{M_eff_Fourier}
\end{subequations}
linking the effective transfer matrix to the effective dielectric tensor. Some comments about equation (\ref{M_eff_expansion}): Similarly to Oldano \cite{Oldano_PRB_96}, we borrowed a quantum mechanics approach, but in our case the perturbation is time-independent, as the grating periodicity is not in the direction of propagation. The three terms appearing in the summation (\ref{M_eff_expansion}) correspond to the 0$^{\rm th}$, 2$^{\rm nd}$ and 3$^{\rm rd}$ order of the perturbation theory corrections, respectively, the 1$^{\rm st}$ order being null. In particular, the 0$^{\rm th}$ order term is the space average of the transfer matrix and N$^{\rm th}$ order corrections describe N-photon scattering, where the coupling between propagating waves is mediated by $(N-1)$ evanescent waves. 
We point out that our procedure gives a double expansion in the scattering multiplicity and in the order of Fourier components. Regarding the short pitch limit, we do not show explicitly the dependence on the grating's pitch and we only present the limit ${\lambda_{\text{g}} \rightarrow 0}$. 

We report explicit expressions for the effective dielectric tensor including the 2-photon scattering terms, and neglecting terms corresponding to scattering multiplicities $N>2$. They have been obtained by calculating $\mathbf{M}^{(\rm{eff})}$ analytically from (\ref{M_Fourier}) and then using equations (\ref{M_eff_Fourier}) to determine the six independent dielectric tensor components. For example, $1/\epsilon^{(\rm{eff})}_{zz}$ can be found from (\ref{M_eff_Fourier_b}), ${\epsilon^{(\rm{eff})}_{z\beta}}/{\epsilon^{(\rm{eff})}_{zz}}$ from (\ref{M_eff_Fourier_a}), so that then $\epsilon_{\alpha \beta}^{(\rm{eff})}$ follows from (\ref{M_eff_Fourier_c}). Recalling that $\mathbf{q}_n=\left\{q_x^{(0)}+n \lambda_0 / \lambda_{\text{g}},q_y^{(0)},0\right\}$, we finally find
\begin{widetext}
\begin{subequations}
\begin{align}
\epsilon^{(\rm{eff, TM})}_{\alpha\beta} & = \xi_{\alpha\beta}^{(0)}-\frac{2}{\left[\beta_{xz}^{(0)}\right]^2+\xi_{xx}^{(0)}\eta_{zz}^{(0)}} \sum_{n=1}^\infty \left( \eta_{zz}^{(0)} \Re \left[ \xi_{x\alpha}^{(n)}\xi_{x\beta}^{(n)*}\right] -\xi_{xx}^{(0)} \Re \left[ \beta_{\alpha z}^{(n)}\beta_{\beta z}^{(n)*}\right] + \right. \\ 
& \left. + \beta_{xz}^{(0)} \Re \left[ \xi_{x\alpha}^{(n)}\beta_{\beta z}^{(n)*}+\xi_{x\beta}^{(n)}\beta_{\alpha z}^{(n)*}\right] \right) \\
\left[\epsilon^{(\rm{eff, TM})}_{zz} \right]^{-1}& =  \eta_{zz}^{(0)}  +\frac{2}{\left[\beta_{xz}^{(0)}\right]^2+\xi_{xx}^{(0)}\eta_{zz}^{(0)}} \sum_{n=1}^\infty \left( \eta_{zz}^{(0)} \left| \beta_{xz}^{(n)}\right|^2 -\xi_{xx}^{(0)} \left| \eta_{zz}^{(n)}\right|^2 -2 \beta_{xz}^{(0)} \Re \left[ \eta_{zz}^{(n)}\beta_{xz}^{(n)*}\right]\right) \\ \nonumber
\frac{\epsilon_{\alpha z}^{(\text{eff, TM})}}{\epsilon^{(\rm{eff, TM})}_{zz}} &=\frac{\epsilon_{z\alpha}^{(\text{eff, TM})}}{\epsilon^{(\rm{eff, TM})}_{zz}}=  \beta_{\alpha z}^{(0)}-\frac{2}{\left[\beta_{xz}^{(0)}\right]^2+\xi_{xx}^{(0)}\eta_{zz}^{(0)}} \sum_{n=1}^\infty \left( \eta_{zz}^{(0)}  \Re \left[ \xi_{x\alpha}^{(n)}\beta_{\alpha z}^{(n)*}\right] +\xi_{xx}^{(0)} \Re \left[ \eta_{zz}^{(n)}\beta_{\alpha z}^{(n)*}\right] +\right.  \\ 
&  \left. + \beta_{xz}^{(0)}\Re \left[ \beta_{xz}^{(n)}\beta_{\alpha z}^{(n)*}- \xi_{x\alpha}^{(n)}\eta_{zz}^{(n)*}\right]\right),
\end{align}
\label{e_eff_transfer_matrix}
\end{subequations}
\end{widetext}
where the superscript TM indicates that the dielectric tensor has been obtained by a transfer matrix approach and $\alpha, \beta \in \{x,y\}$. We point out that retaining only the first term in the r.h.s. of each of the equations above corresponds to the space average approximation (labeled TM00 in section V), while the summations correspond to the 2-photon scattering correction expressed as a Fourier expansion and are included in the approximations labeled TM21 and TM22 in section V, where the former includes only Fourier terms with $n=1$ and the latter terms with $n\le 2$. The above formulas allow the calculation of an effective macroscopic tensor when the explicit form of the periodic mesoscopic dielectric tensor is specified. Equations (\ref{e_eff_transfer_matrix}), together with the transfer matrix expansion (\ref{M_eff_expansion}), are the main results of this section.
In Section IV we specify the mesoscopic dielectric tensor for the case of DHFLCs and, in the Appendix, we explicitly report the corresponding Fourier components, up to the second order, of the coefficients appearing in (\ref{e_eff_transfer_matrix}).

\section{\label{sec:level1}Bloch wave approach}
In this section, we adopt the Bloch wave method described by Galatola \cite{Galatola_97} and extended  by Ponti et al. \cite{Ponti_PRE_2001}. We will focus on the short-pitch limit, {\em i.e.} $\lambda_{\text{g}}/\lambda_0 \rightarrow 0$, thus neglecting all terms proportional to $(\lambda_{\text{g}}/\lambda_0)^m$ with $m>0$. Including terms corresponding to the 0-, 2- and 3-photon scattering, as in the previous section, we write the effective dielectric tensor as
\begin{widetext}
\begin{equation}\label{Bloch_e_eff_expansion}
\boldsymbol{\epsilon}^{(\text{eff,BW})} = \boldsymbol{\epsilon}^{(0)}+\sum_{n\neq 0}\boldsymbol{\epsilon}^{(n)}\cdot \boldsymbol{G}^{\rm (0)} \cdot \boldsymbol{\epsilon}^{(-n)} + \sum_{n \ne 0}\sum_{m \ne \{0,n\}} \boldsymbol{\epsilon}^{(n)}\cdot \boldsymbol{G}^{\rm (0)} \cdot \boldsymbol{\epsilon}^{(m)}\cdot \boldsymbol{G}^{\rm (0)} \cdot \boldsymbol{\epsilon}^{(-n-m)},
\end{equation}
\end{widetext}
where the superscript BW indicates that the above formula has been derived using a Bloch wave approach, 
\begin{equation}
\boldsymbol{\epsilon}^{(n)}  = \frac{1}{\lambda_{\text{g}}} \int_0^{\lambda_{\text{g}}} e^{-i n k_{\text{g}} x}\boldsymbol{\epsilon} \; {\rm d}x,
\label{Fourier_coefficient_G}
\end{equation}
and $\boldsymbol{G}^{\rm (0)}$ is the short pitch limit of the matrix defined in equation (4) of ref. \cite{Ponti_PRE_2001}, given explicitly for our orientation by
\begin{equation}
\boldsymbol{G}^{\rm (0)}=-\frac{1}{\epsilon^{(0)}_{xx}}\begin{pmatrix} 1 &  0 &  0 \\   0 &  0 & 0 \\  0 & 0 & 0\end{pmatrix}.
\end{equation}
We note that $\boldsymbol{\epsilon}^{(0)}$ is simply the space average of the dielectric tensor, while the other terms represent corrections due to the coupling with evanescent waves mediated by the grating. We will show in section V that for our purposes the 3-photon scattering terms can be neglected. Using the above definitions and including only 0- and 2-photon scattering terms (as done in the previous section), we can calculate the components of the effective dielectric tensor explicitly as 
\begin{subequations}\label{e_eff_Bloch_def}
\begin{align}
\epsilon_{\alpha\beta}^{(\text{eff,BW})} &= \epsilon_{\alpha\beta}^{(0)} -\frac{2}{\epsilon_{xx}^{(0)}} \sum_{n=1}^\infty \Re\left[\epsilon_{x\alpha}^{(n)}\epsilon_{x\beta}^{(n)*}\right], \\
\epsilon_{\alpha z}^{(\text{eff,BW})} &=\epsilon_{z\alpha}^{(\text{eff,BW})}=\epsilon_{\alpha z}^{(0)} -\frac{2}{\epsilon_{xx}^{(0)}} \sum_{n=1}^\infty \Re\left[\epsilon_{x\alpha}^{(n)}\epsilon_{xz}^{(n)*}\right],\\
\epsilon_{zz}^{(\text{eff,BW})} &= \epsilon_{xx}^{(0)} +\frac{2}{\epsilon_{xx}^{(0)}} \sum_{n=1}^\infty \left|\epsilon_{xz}^{(n)}\right|^2.
\end{align}
\end{subequations}
where the superscript BW indicates that the dielectric tensor has been obtained by a Bloch wave approach and, again, $\alpha, \beta \in \{x,y\}$.  We note that the series expansion (\ref{Bloch_e_eff_expansion}) is formally very similar to the  transfer matrix expansion (\ref{M_eff_expansion}) and both describe the same physical processes. However, the various terms are different, including the 0$^{\rm th}$ order. Again, we note that retaining only the first term in the r.h.s. of each of the equations above corresponds to the space average approximation (labeled BW00 in section V), while the summations correspond to the 2-photon scattering correction expressed as a Fourier expansion. We add that these summations are included in the approximations labeled BW21 and BW22 in section V, where the former includes only Fourier terms with $n=1$ and the latter terms with $n\le 2$.
We will see in the next section that the Bloch wave method produces more convenient analytical formulas, when applied to DHFLCs, compared to the transfer matrix approach. We will compare the accuracy of the two methods in section V, but we can anticipate that the two methods give similar accuracy. 
\section{\label{sec:level1}Effective dielectric tensor for short-pitch deformed helix ferroelectric liquid crystals} 
\begin{figure}[h]
\includegraphics[width=15cm]{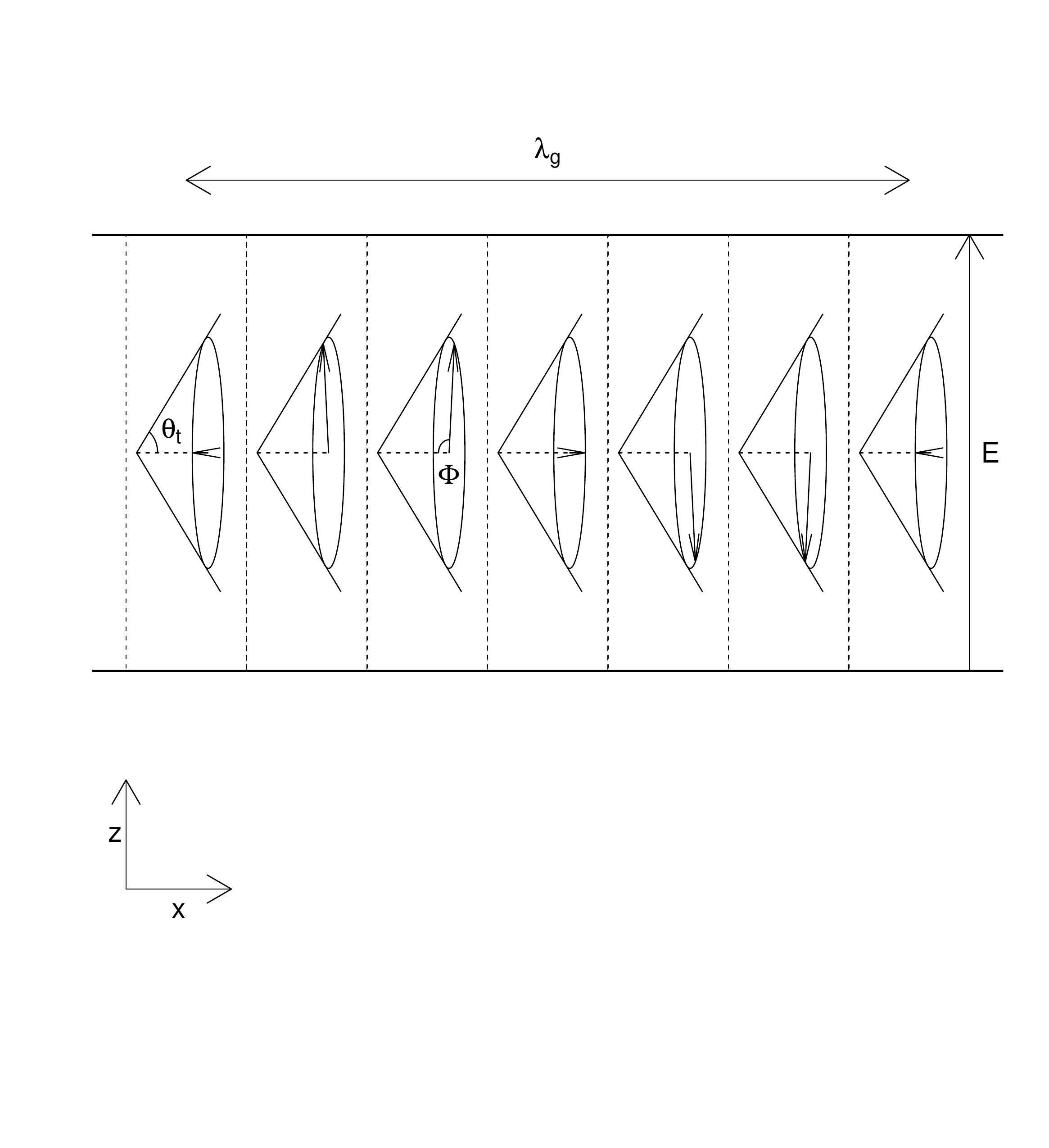}
\caption{Geometry of a DHFLC. In each smectic layer (delimited by dashed lines) the direction of the director is shown by an arrow. The resulting helical structure, with pitch $\lambda_{\text{g}}$, has its axis along $x$. The tilt angle, $\theta_t$, and the azimuthal angle, $\Phi$, are also indicated. The electric field, $\mathbf{E}$, is perpendicular to the helix axis.}
\label{Fig_helix}
\end{figure}

In this section, we explicitly calculate the effective dielectric tensor of deformed helix ferroelectric liquid crystals, assuming the same slab geometry of Figure \ref{Fig_geometry}. These LCs are used in their Smectic C* phase with homogeneous alignment under small electric fields, which produce a deformation of the helical structure rather than its complete unwinding \cite{Beresnev_LC_1989}. Under these assumptions the mesoscopic dielectric tensor takes the form \cite{Kiselev2011}
\begin{align}
\epsilon_{ij} = \epsilon_{\bot}\delta_{ij}+\delta\epsilon\;\*d_{i}\*d_{j},
\label{tensor}
\end{align}
where $i,j \in \{x,y,z\}$, $\delta_{ij}$ is the Kronecker delta, $\epsilon_{\bot}$ ($\epsilon_{\parallel}$) is the relative permittivity corresponding to the ordinary (extraordinary) refractive index, and $\delta\epsilon =\epsilon_{\parallel}-\epsilon_{\bot}$. The director, $\uv{d}$, is a unit vector giving the orientation of all molecules in each layer. Choosing the helix axis along $x$ we have, 
\begin{align}
\uv{d} = (d_{x}, d_{y}, d_{z}) = (\cos{\theta_{\text{t}}}, \ \sin{\theta_{\text{t}}}\*\cos{\Phi}, \   \sin{\theta_{\text{t}}}\*\sin{\Phi}),
\label{director}
\end{align}
where $\theta_{\text{t}}$ is the tilt angle defining the director cone, and $\Phi$ is the azimuthal angle made by the director on the cone about the helix axis in a given smectic layer. We now consider small electric fields applied perpendicular to the helix axis and to the slab surface, i.e. $\mathbf{E}=E\*\uv{z}$, with $|\mathbf{E}| = E \ll E_{C}$. This geometry is illustrated in Figure \ref{Fig_helix}. For convenience, we also define the dimensionless small parameter $\alpha_{E} \equiv E/E_{C} \ll 1$. Under the above assumptions, the deformed helical structure is given by \cite{Abdulhalim1991,Kiselev2011}
\begin{equation}\label{Phi}
\Phi \approx \phi(x) + \alpha_{E}\*\sin{\phi(x)},
\end{equation}
where $\phi(x) = k_{\text{g}}\*x = 2\*\pi\*x/\lambda_{\text{g}}$.
We can now explicitly calculate the Fourier coefficients presented in the previous sections. Since the above formula is only valid for small electric fields, we expand the analytical formulas in this section up to the second order in the Taylor expansion of the small parameter $\alpha_{E}$.

We report here the results obtained with the Bloch wave method, while analytical formulas for the transfer matrix approach are presented in the Appendix. The effective dielectric tensor can be readily obtained from equations (\ref{e_eff_Bloch_def}), using definitions (\ref{director}), (\ref{tensor}) and (\ref{Fourier_coefficient_G}). Including terms up to the Fourier coefficient $n=2$, we get
\begin{subequations}
\begin{align}
\epsilon_{xx}^{(\text{eff})} &\approx \epsilon_\bot + \delta\epsilon \cos^2\theta_t \label{xxeff}, \\
\epsilon_{xy}^{(\text{eff})} &= \epsilon_{yx}^{(\text{eff})} \approx - \alpha_E \frac{\delta\epsilon}{2} \sin\theta_t \cos\theta_t,  \label{xyeff} \\
\epsilon_{yy}^{(\text{eff})} &\approx \frac{\epsilon_\bot}{2}\left(1+\frac{\epsilon_\parallel}{\epsilon_{xx}^{(\text{eff})}} \right)+\alpha_E^2\frac{\delta\epsilon}{4}\sin^2\theta_t, \label{yyeff}  \\
\epsilon_{xz}^{(\text{eff})} &=\epsilon_{zx}^{(\text{eff})}=\epsilon_{yz}^{(\text{eff})} =\epsilon_{zy}^{(\text{eff})}=0, \\
\epsilon_{zz}^{(\text{eff})} &\approx \frac{\epsilon_\bot}{2}\left(1+\frac{\epsilon_\parallel}{\epsilon_{xx}^{(\text{eff})}} \right)-\alpha_E^2\frac{\delta\epsilon}{4}\frac{\epsilon_\bot }{\epsilon_{xx}^{(\text{eff})}}\sin^2\theta_t,
\end{align}
\label{e_eff_DHFLC}
\end{subequations}
corresponding to the BW22 approximation of section V. 
It is also useful to derive explicit expressions for the eigenvalues of the effective dielectric tensor, the rotation of the principal axes and the birefringence. The three eigenvalues are $\epsilon_{+}^{(\text{eff})}$, $\epsilon_{-}^{(\text{eff})}$ and $\epsilon_{zz}^{(\text{eff})}$, with
\begin{equation}
\epsilon_{\pm}^{(\text{eff})}=\left[\frac{\epsilon_{xx}^{(\text{eff})}+\epsilon_{yy}^{(\text{eff})}}{2}\right]\pm\left[\frac{\epsilon_{xx}^{(\text{eff})}-\epsilon_{yy}^{(\text{eff})}}{2}\right]\sqrt{1+\tan^2\left[2\Omega\right]},
\end{equation}
and the rotation of the principal axes around $z$, $\Omega$, induced by the electric field is defined by
\begin{align}
\tan{(2\*\Omega)} = \frac{2 \epsilon_{xy}^{\text{(eff)}}}{\epsilon_{xx}^{\text{(eff)}}-\epsilon_{yy}^{\text{(eff)}}}.
\end{align}
From the above expression we can explicitly find
\begin{align}
\Omega = -\frac{\alpha_E}{4}\frac{\delta\epsilon \sin(2\theta_t)}{\epsilon_{xx}^{\text{(eff)}}-\epsilon_{yy,0}^{\text{(eff)}}}
\end{align}
where we defined $\epsilon_{yy,0}^{\text{(eff)}}=\frac{\epsilon_\bot}{2}\left(1+\frac{\epsilon_\parallel}{\epsilon_{xx}^{(\text{eff})}} \right)$.
Finally, the birefringence is given by
\begin{widetext}
\begin{equation}
\Delta n \equiv \sqrt{\epsilon_{+}^{(\text{eff})}}-\sqrt{\epsilon_{-}^{(\text{eff})}}=\sqrt{\epsilon_{xx}^{(\text{eff})}}-\sqrt{\epsilon_{yy,0}^{\text{(eff)}}}+\alpha_E^2 \frac{\sqrt{\epsilon_{xx}^{(\text{eff})}}\sqrt{\epsilon_{yy,0}^{\text{(eff)}}}-\epsilon_\bot}{\epsilon_{xx}^{(\text{eff})}\sqrt{\epsilon_{yy,0}^{\text{(eff)}}}-\epsilon_{yy,0}^{\text{(eff)}}\sqrt{\epsilon_{xx}^{(\text{eff})}}}\frac{\delta\epsilon\sin^2\theta_t}{8}.
\end{equation}
\end{widetext}
The formulas in this section can be used to describe DHFLCs using an effective dielectric tensor approach and we will show in the next section that they are remarkably accurate within a wide range of parameters. We notice that in the absence of any electric field, {\em i.e.} when $\alpha_E$=0, equations (\ref{e_eff_DHFLC}) agree with the results in ref.\cite{Oldano_PRB_96} (cfr. equation 9).

\section{Exact numerical results and discussion}
In this section we benchmark the effective dielectric tensor approximations derived in the previous sections against exact numerical calculations for a DHFLC slab in the presence of an external electric field. In particular, we compute the birefringence, $\Delta n$, the rotation of the optic axes induced by the electric field, $\Omega$, and the transmission at crossed polarizers, adopting an approach similar to the one used in \cite{Galatola_97}. In the case of the exact numerical calculations, the first step of the procedure consists in constructing the block matrix $\boldsymbol{\mathcal{M}}$ including all the matrices $\mathbf{M}_{n m}$ with $\{n,m\} \le n_{\rm max}$ and $|n-m|\le n_{\rm max}$. We found that $n_{\rm max}=4$ is sufficient to obtain a typical accuracy of 0.001\% on the calculated optical properties. From the eigenvalues and eigenvectors of the transfer matrix at normal incidence we can immediately calculate the birefringence and the rotation of the principal axes around $z$, respectively, while transmission at any incident angle can be calculated using the procedure described in detail in ref.\cite{Galatola_97}. In the case of the various effective medium approximations, we adopt the same approach, but using a 4$\times$4 transfer matrix calculated with equations (\ref{M_eff_Fourier}) from the various effective dielectric tensors.
The LC parameters are those of the mixture FLC-576 at room temperature, {\em i.e.} $\theta_t$=32$^\circ$,  $\epsilon_\bot$=1.5, $\epsilon_\parallel$=1.72 and  $\lambda_{\text{g}}$=0.2 $\mu$m \cite{Kiselev2011}. The reference wavelength of the incident light has been chosen as $\lambda_0=$1.55 $\mu$m, as this is the typical telecommunication wavelength we use in our optical transducers \cite{brodzeli2013sensors}. Unless otherwise specified, other parameters are $d=50$ $\mu$m, $\theta_i$=0, $\varphi_i$=0, $\alpha_E$=0.2 and $n_{\text{m}}$=1.5. For each approach, we have chosen to include in the comparison the space average approximation, {\em i.e.} the 0$^{\rm th}$ order in the scattering multiplicity, and the 2-photon scattering approximation, the latter with $|n|\le 1$ or $|n|\le 2$ in the Fourier expansion. This gives a total of 6 approximations, labeled TM00, TM21, TM22, BW00, BW21, BW22, where TM (BW) indicates the transfer matrix (Bloch wave) method and the two numbers indicate the largest multiplicity of the photon scattering and the largest Fourier component included, respectively. A summary of the various approximations is presented in Table \ref{Table_app}.
\begin{table}[h!]
\begin{center}
\begin{tabular}{ c | c | l |  l | l}
 Approximation  & Equation & R.h.s terms included & Plot legend & Notes\\ \hline
TM00 & \ref{e_eff_transfer_matrix} & 1$^{st}$ & Dashed green & Same as ref.\cite{Kiselev2011} \\ 
TM21 & \ref{e_eff_transfer_matrix} & 1$^{st}$, $\sum_{n \le 1}$ & Dashed red & See Appendix \\
TM22 & \ref{e_eff_transfer_matrix} & 1$^{st}$, $\sum_{n \le 2}$ & Dashed blue &  See Appendix \\
BW00 & \ref{e_eff_Bloch_def} &  1$^{st}$  & Solid green &  \\
BW21 & \ref{e_eff_Bloch_def} & 1$^{st}$, $\sum_{n \le 1}$& Solid red & \\
BW22 & \ref{e_eff_Bloch_def} & 1$^{st}$, $\sum_{n \le 2}$ & Solid blue & See eq.(\ref{e_eff_DHFLC}) \\ 
Exact & N/A & N/A & black crosses & See ref.\cite{Galatola_97} \\ \hline
\end{tabular}
\caption{List of approximations compared in this section. The second and third column report the formulas used to calculate the effective dielectric tensor and the terms included, respectively. Including only the 1$^{st}$ term corresponds to the space average approximation. The fourth column indicates the lines and colors used in the Figures to indicate each approximation. Analytical expressions for BW00 and BW21 are not explicitly reported in this paper as they can be easily derived.}
\label{Table_app}
\end{center}
\end{table}

\begin{figure}[h]
\includegraphics[width=15cm]{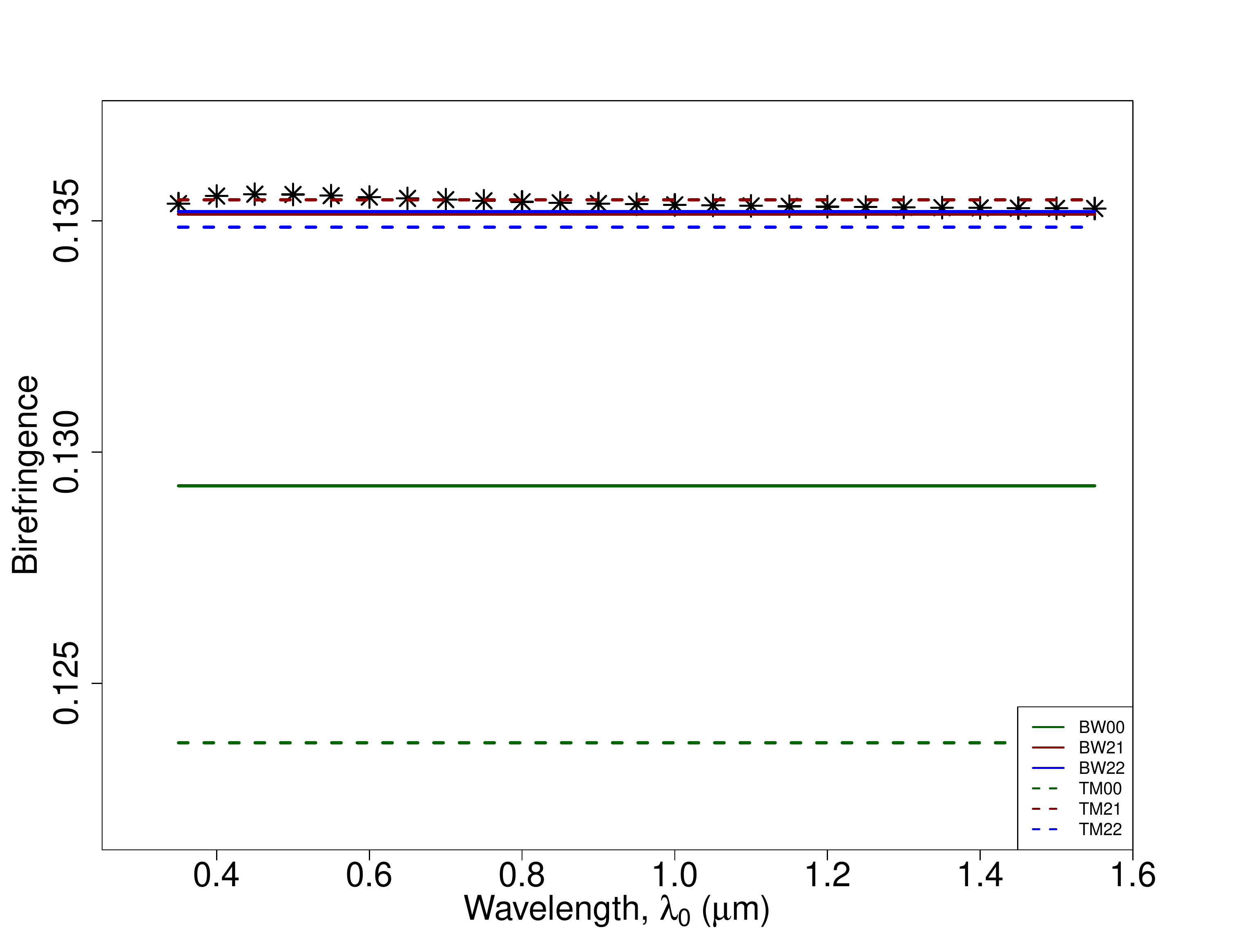}
\caption{Birefringence as a function of the incident light's wavelength ($\lambda_0$) for the exact numerical calculations (black crosses) and for the 6 analytical approximations, as indicated in the legend. A summary of the various approximations is also reported in Table \ref{Table_app} for convenience.}
\label{Fig_dn_lambda_0}
\end{figure}
\begin{figure}[h]
\includegraphics[width=15cm]{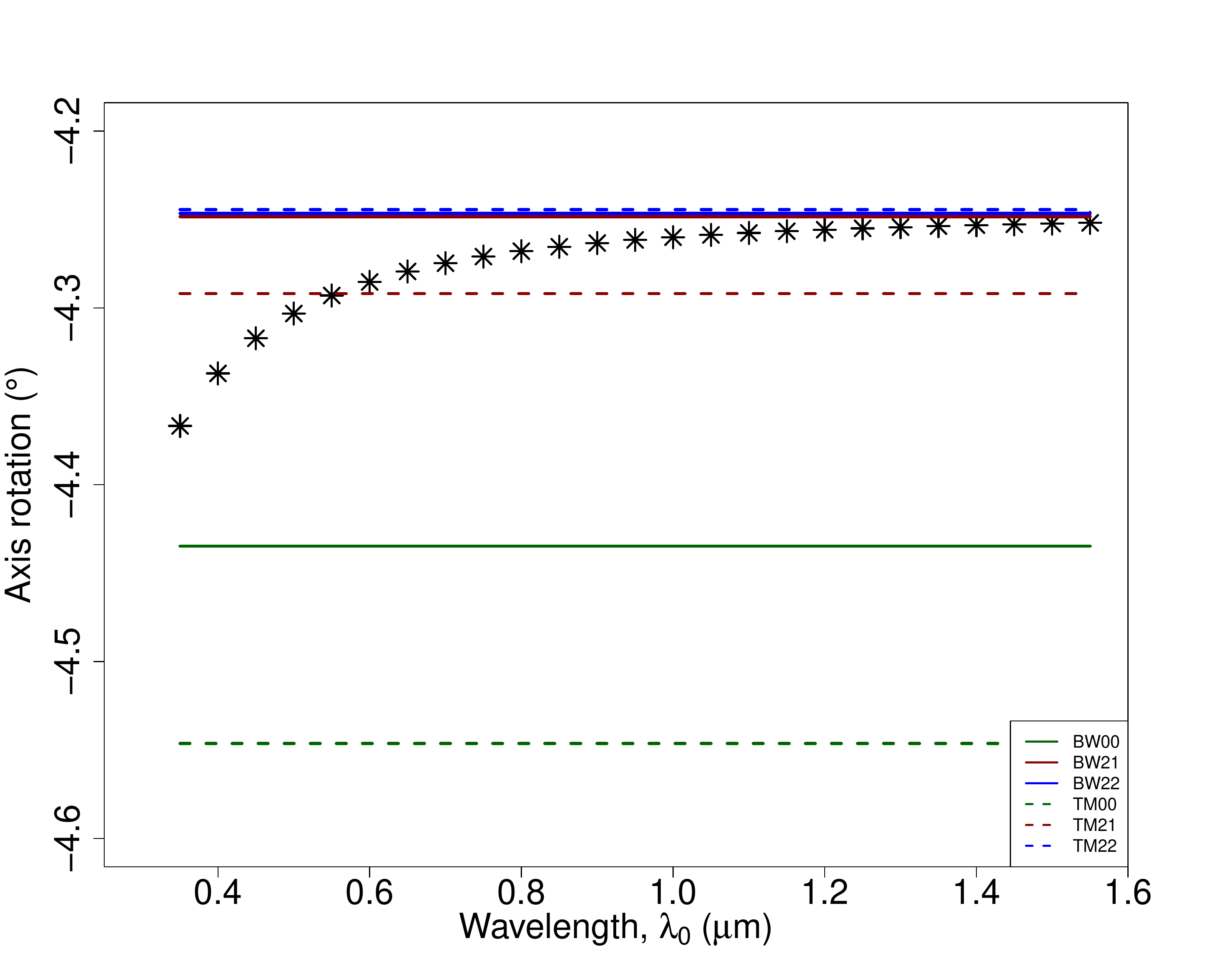}
\caption{Rotation of the optic axes as a function of the incident light's wavelength $\lambda_0$. Symbols indicate the exact numerical calculations, while lines denote the 6 analytical approximations, as indicated in the legend.}
\label{Fig_omega_lambda_0}
\end{figure}

In Figures \ref{Fig_dn_lambda_0} and \ref{Fig_omega_lambda_0} we show the birefringence, $\Delta n$, and the optic axes rotation, $\Omega$, respectively, in the presence of an electric field ($\alpha_{E}$=0.2), as a function of the incident wavelength, $\lambda_0$. These calculations allow us to check the validity of the short pitch approximation. For $\Delta n$, we see that at long wavelengths the exact results tend to the values calculated with the 2-photon scattering approximations, namely TM21, TM22, BW21, BW22. The various approximations predict the short pitch limit with different relative accuracies, all below 0.3\%, with the Bloch wave method giving the best results. It is interesting to note that the exact birefringence is within 0.3\% of its short pitch limit already for $\lambda_0$=0.35 $\mu$m, corresponding to $\lambda_{\text{g}}/\lambda_0$=0.57. For the optic axes rotation the results are similar, except that the TM22 approximation gives much better results than in the birefringence case, as opposed to the TM21 which performs poorly. For the optic axes rotation, the exact results are within within 1\% of the short pitch limit for $\lambda_0$=0.55 $\mu$m (corresponding to $\lambda_{\text{g}}/\lambda_0$=0.36) and within 0.3\% of the same limit for $\lambda_0 >$ 0.9 $\mu$m. We conclude that the effective medium approximation is appropriate for our parameters and it is an excellent approximation even at much shorter wavelengths. 
\begin{figure}[h]
\includegraphics[width=15cm]{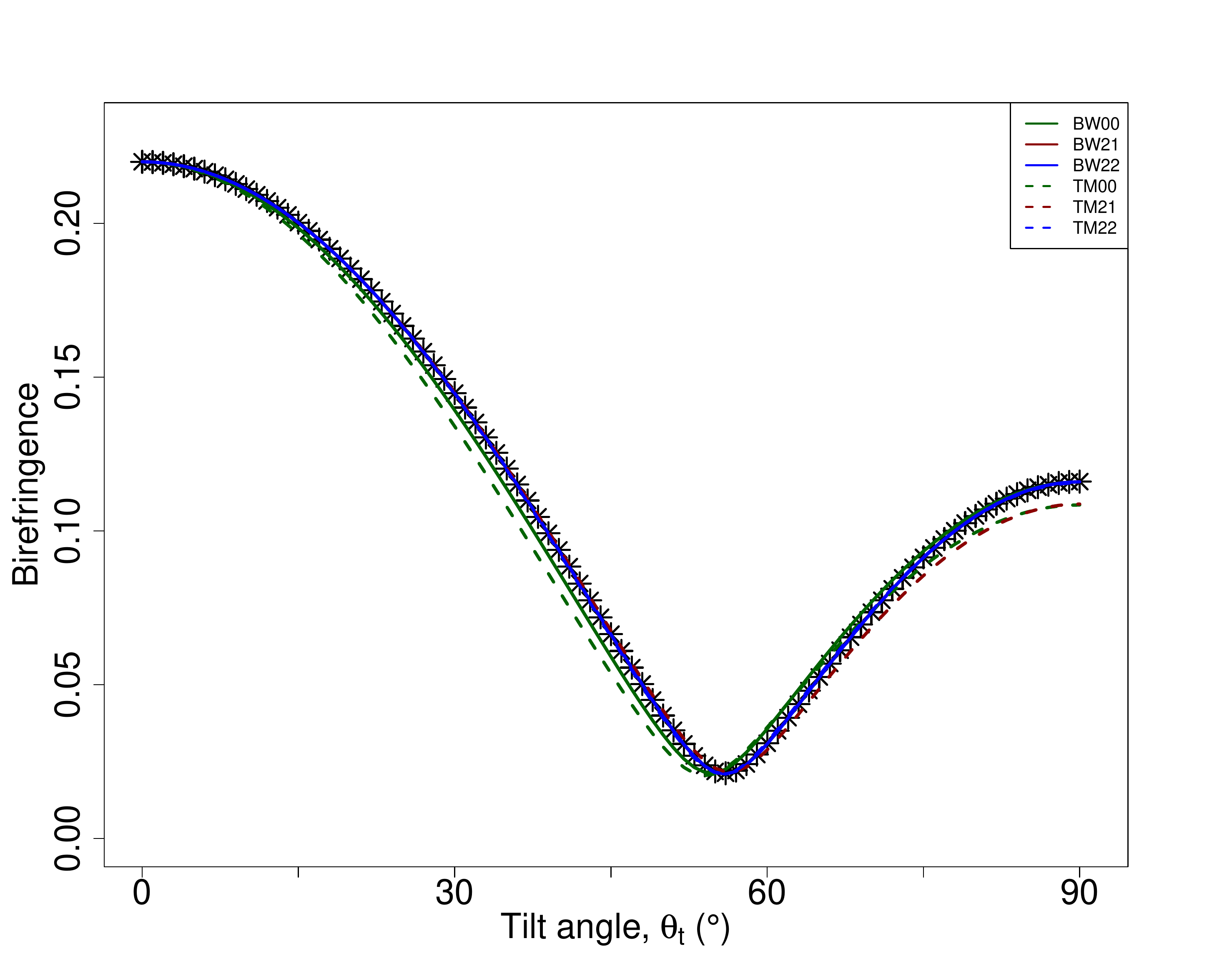}
\caption{Birefringence as a function of the tilt angle. Symbols indicate the exact numerical calculations, while lines denote the 6 analytical approximations, as indicated in the legend.}
\label{Fig_dn_tilt}
\end{figure}
\begin{figure}[h]
\includegraphics[width=15cm]{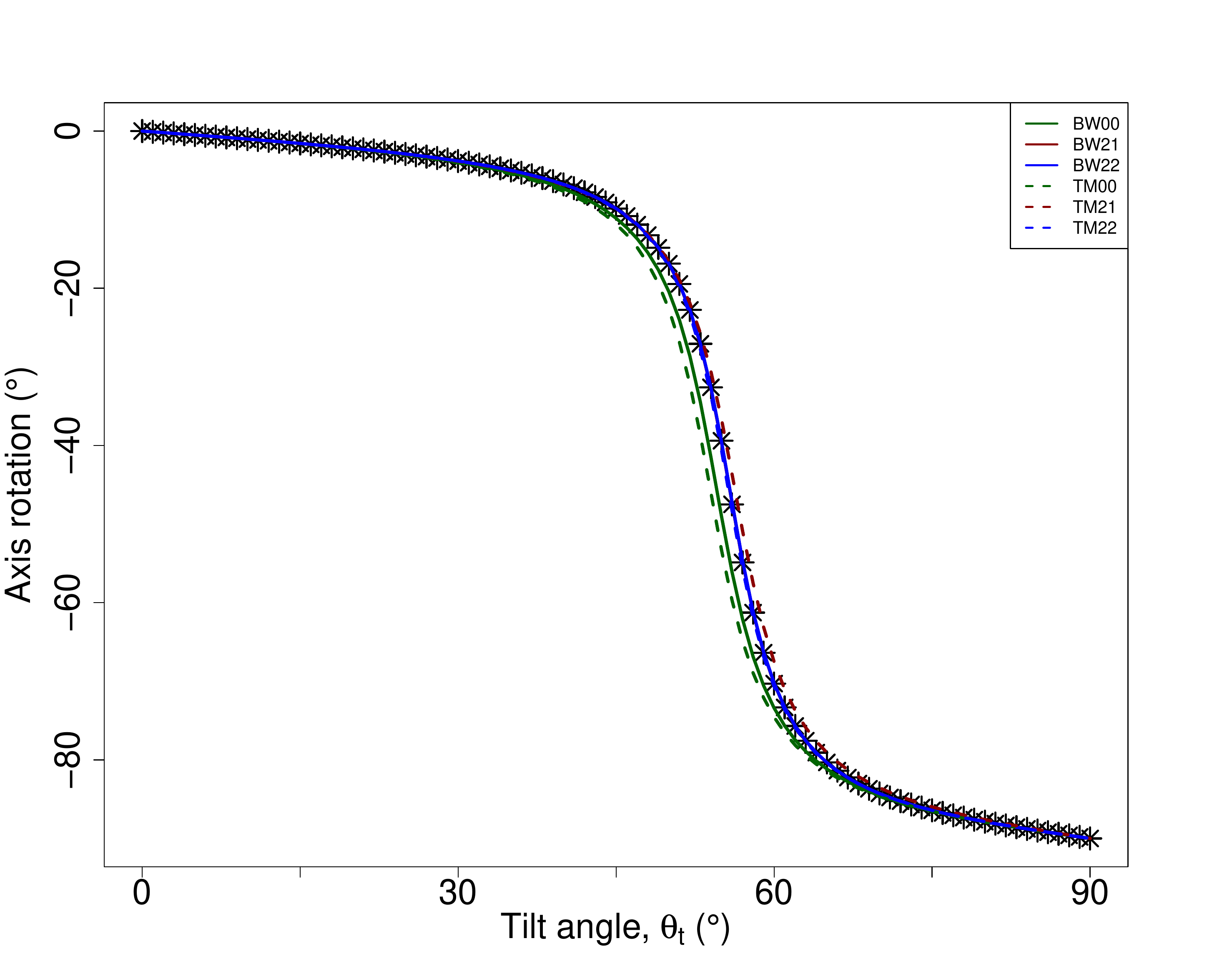}
\caption{Rotation of the optic axes as a function of the tilt angle. Symbols indicate the exact numerical calculations, while lines denote the 6 analytical approximations, as indicated in the legend.}
\label{Fig_omega_tilt}
\end{figure}
In Figures \ref{Fig_dn_tilt} and \ref{Fig_omega_tilt} we show the birefringence, $\Delta n$, and the optic axes rotation, $\Omega$, respectively, in the presence of an electric field ($\alpha_{E}$=0.2) as a function of the tilt angle. The birefringence shows the expected behavior, reaching a minimum close to the isotropization angle, but the medium never becomes isotropic due to the presence of an electric field \cite{Kiselev_2014}. We see that for tilt angles larger than 15$^\circ$ the space average approximation gives a noticeable error on the birefringence. Moreover, neither the TM00 nor the TM21 approximation gives the correct limit for cholesteric LCs ($\theta_t$=90$^\circ$).  In the case of the axis rotation the discrepancy between the space average approximations and the exact results is noticeable at tilt angles between $30^\circ$ and $60^\circ$, while the correct behavior is recovered in the limit of cholesteric LCs. We conclude that, for tilt angles $>15^\circ$, 2-photon scattering terms should be included. The actual error committed in calculating the optical transmission (or reflection) of a DHFLC cell depends on the cell's thickness.
\begin{figure}[h]
\includegraphics[width=15cm]{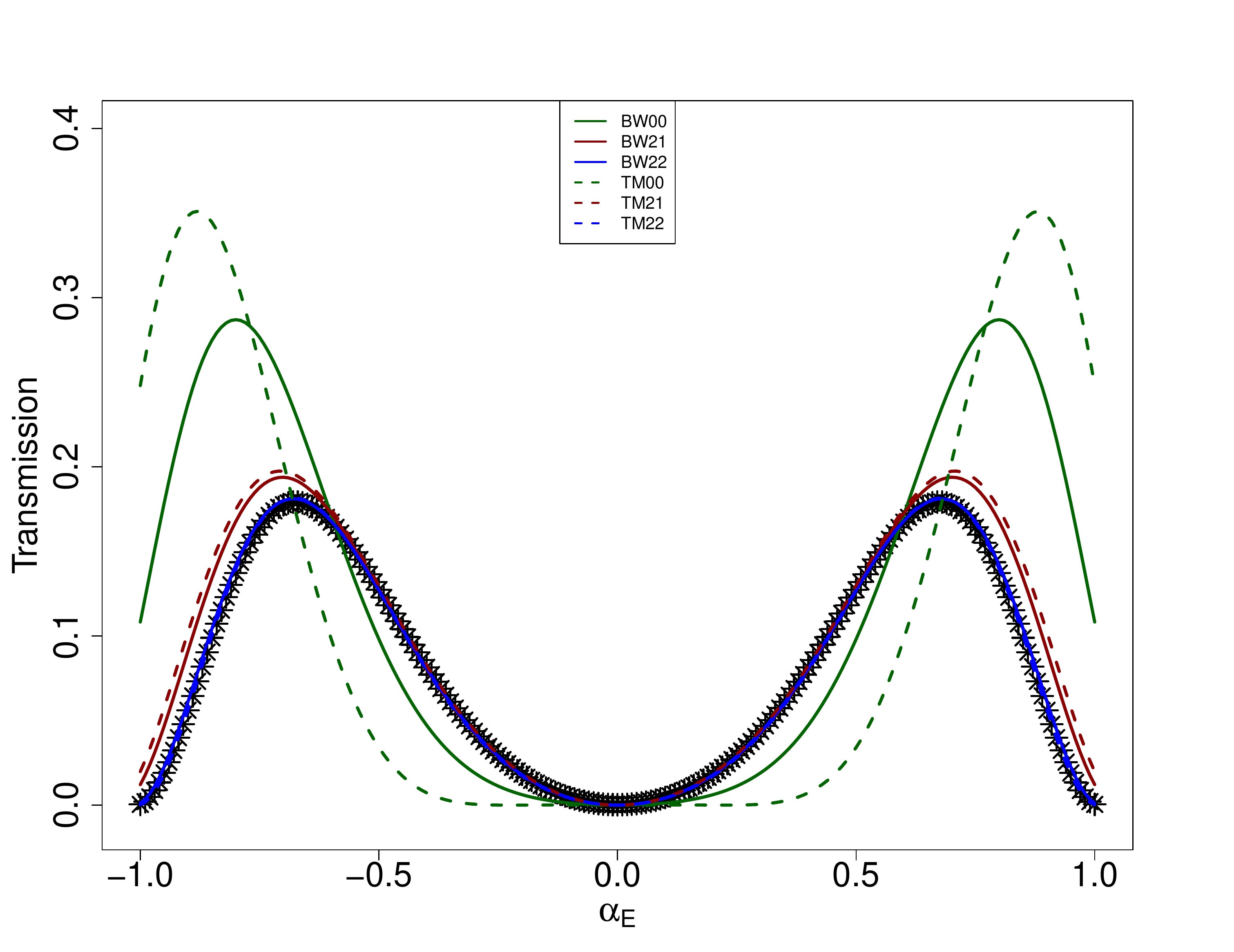}
\caption{Normal incidence transmission at crossed polarisers calculated as a function of the electric field. Incident light is linearly polarised along the direction of the helix axis. Symbols indicate the exact numerical calculations, while lines denote the 6 analytical approximations, as indicated in the legend.}
\label{Fig_alpha}
\end{figure}
In order to further prove this point, we show in Figure \ref{Fig_alpha} an example of normal incidence transmission at crossed polarizers, when the incident light is polarized along the direction of the helix axis and $\theta_t=32^\circ$. We notice that the approximations with 2-photon and first order Fourier components (BW21, TM21) give already very good results, at least within the range of validity of the small field approximation. The more accurate approximations, BW22 and TM22, give an almost perfect agreement up to $\alpha_E$=1. The space average approximations instead (BW00 and TM00) are in poor agreement with the exact results at all electric field intensities. 

\begin{figure}[h]
\includegraphics[width=15cm]{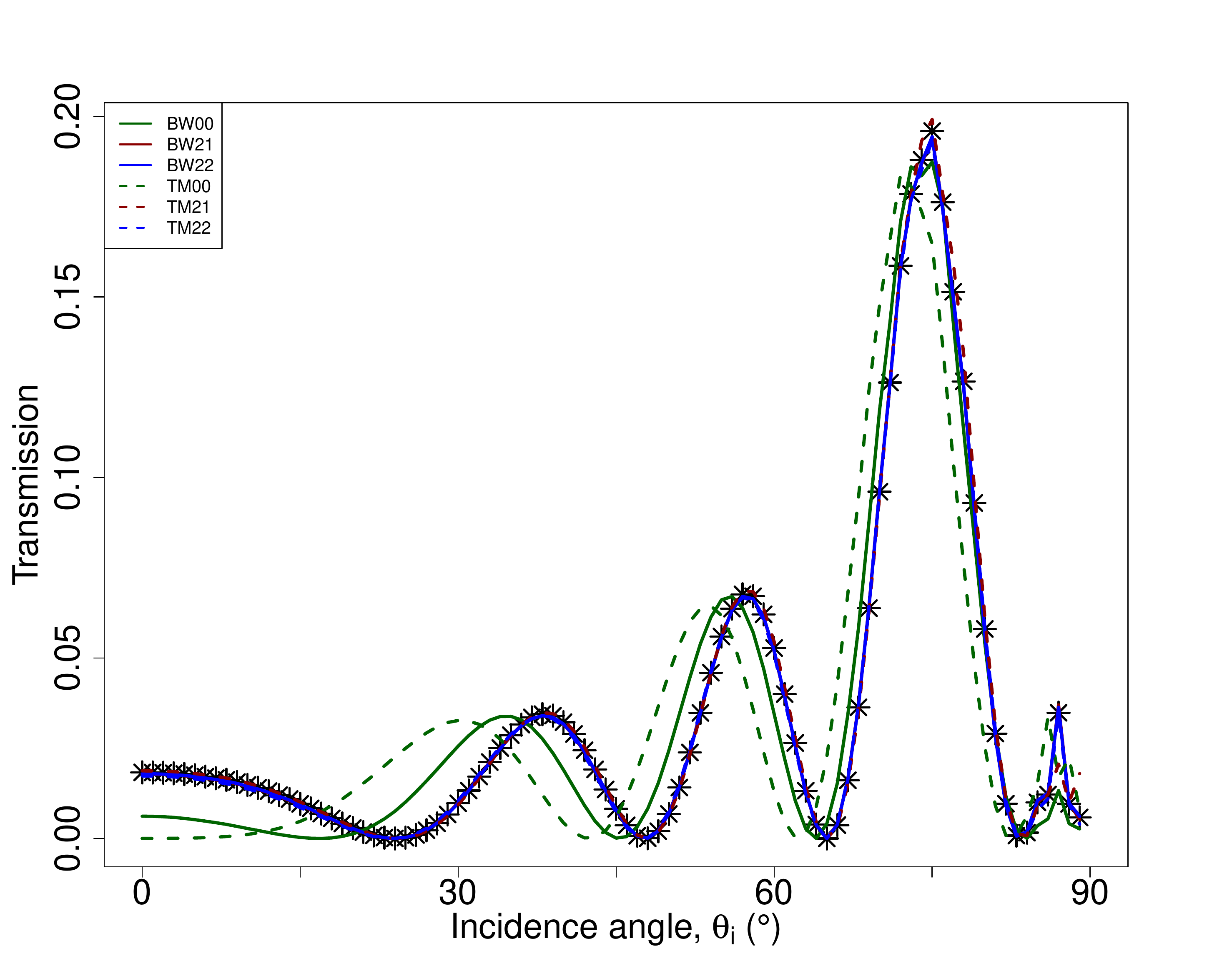}
\caption{Transmission at crossed polarisers as a function of the incident angle. Incident light is linearly polarised along the direction of the helix axis. Symbols indicate the exact numerical calculations, while lines denote the 6 analytical approximations, as indicated in the legend.}
\label{Fig_theta_i}
\end{figure}
Since the $zz$ component of the effective dielectric tensor can only be probed at oblique incidence, we have calculated the transmission at crossed polarisers for various incident angles in the presence of an electric field ($\alpha_E$=0.2). The results, reported in Figure \ref{Fig_theta_i}, show that the 2-photon approximation works well in the whole range of incident angles, while the space average approximations consistently produce noticeable errors.

In general, we conclude that, while space average approximations are appropriate for small tilt angles, 2-photon scattering terms must be taken into account for DHFLCs with large tilt angles. We have also found that first order Fourier components are often enough to get good results across a wide range of parameters, but we recommend including second order Fourier terms, which make the agreement with the exact calculations almost perfect. Our results also confirm that 3-photon scattering terms can be neglected for most practical purposes. 
Another interesting conclusion is that both the transfer matrix and the Bloch wave method have the same accuracy when similar terms are taken into account in the effective dielectric tensor expansion. 
Finally, and most importantly, we have proven that the effective dielectric tensor description is remarkably accurate over a huge range of parameters, covering virtually all practical situations. 
Based on the results of this section, we recommend researchers to use formulas (\ref{e_eff_DHFLC}), which are extremely convenient and remarkably accurate.

\section{Conclusions}
We have presented analytical formulas to calculate the effective dielectric tensor of short pitch DHFLCs with homogeneous alignment in the presence of an external electric field. This effective medium approximation is shown to be extremely accurate in the whole range of practically relevant parameters. We also compared two complementary approaches to the effective medium approximation, relying on an expansion of the mesoscopic transfer matrix and the mesoscopic dielectric tensor, respectively. In order to do that, we have derived for the first time an explicit expansion for the effective transfer matrix and calculated the corresponding effective dielectric tensor. Our results show that the two methods give similar results when terms describing similar physical processes are taken into account.

\acknowledgments
This study was supported by funding from the Australian Research Council’s Discovery Program (grant no. DP160104625).

\appendix*
\section{Transfer matrix method - Coefficients of the effective dielectric tensor expansion for DHFLCs}

In this Appendix, we present analytical expressions to calculate the effective dielectric tensor with the transfer matrix method, using equations (\ref{e_eff_transfer_matrix}). We assume that the periodic material is a DHFLC with the geometry of Figure \ref{Fig_helix} in the presence of small electric fields, so that definitions (\ref{tensor})-(\ref{Phi}) hold. As discussed in Section IV, we expand each coefficient into a Taylor series in the small parameter $\alpha_E$ up to the second order. For the space average approximation, TM00, we only include the 0$^{\rm th}$ order term, i.e. the first term on the r.h.s. of each equation. The relevant quantities are: 
\begin{subequations}
\begin{align}
\beta_{\alpha z}^{(0)} &= \beta_{z \alpha}^{(0)} = 0 \\
\eta_{zz}^{(0)} &\approx \frac{1+v\*\gamma_{v}^{2}\*\alpha_{E}^{2}}{\epsilon_{\bot}\sqrt{1+v}}\\
\xi_{xx}^{(0)} &\approx \xi_{xx,0}^{(0)} + \gamma_{xx}\*\alpha_{E}^{2} \\
\xi_{xy}^{(0)} &= \xi_{yx}^{(0)} \approx \gamma_{xy}\*\alpha_{E} \\
\xi_{yy}^{(0)} &\approx \xi_{yy,0}^{(0)} + \gamma_{yy}\*\alpha_{E}^{2},
\end{align}
\label{oldparam1}
\end{subequations}
where 
 \begin{subequations}
 \begin{align}
 v &= \frac{\delta\epsilon}{\epsilon_\bot} \sin^{2}{\theta_{\text{t}}} \\
 \gamma_{v} &= (\sqrt{1+v}-1)/v \\
 \gamma_{xx} &= \frac{\delta\epsilon \cos^{2}{\theta_{\text{t}}}}{\sqrt{1+v}}\* \ v\*\gamma_{v}^{2} \\ 
 \gamma_{xy} &= -\delta\epsilon\;\gamma_{v}\*\sin{\theta_{\text{t}}}\*\cos{\theta_{\text{t}}} \\
 \gamma_{yy} &= \epsilon_{\bot}\*\sqrt{1+v}\* \ v\*\gamma_{v}^{2}, \\
  \xi_{xx,0}^{(0)} &= \epsilon_{\bot}+ \frac{\delta\epsilon \cos^{2}{\theta_{\text{t}}}}{\sqrt{1+v}},\\
  \xi_{yy,0}^{(0)} &= \epsilon_{\bot}\sqrt{1+v}.
 \end{align}
 \label{oldparam2}
 \end{subequations}
These formulas are identical to the ones presented in ref.\cite{Kiselev2011}, even if the notation is slightly different. For the TM21 and TM22 approximations, we also need the 1$^{\rm st}$ and 2$^{\rm nd}$ order Fourier components, which are as follows: 
 \begin{subequations}
\begin{align}
 \beta_{zx}^{(1)} &= \beta_{xz}^{(1)} \approx \frac{i\*\gamma_{xy}}{\epsilon_\perp\sqrt{1+v}}\*\bigg[1+\frac{3}{4}\*(1-3\*\gamma_{v})\*\alpha_{E}^{2}\bigg] \\
\beta_{zy}^{(1)} &= \beta_{yz}^{(1)} \approx i\*\gamma_{v}^{2}\*v\*\alpha_{E} \\
\eta_{zz}^{(1)} &\approx -\frac{1}{\epsilon_{\bot}}\*\frac{v\*\gamma_{v}^{2}\*\alpha_{E}}{\sqrt{1+v}} \\
\xi_{xx}^{(1)} &\approx -\gamma_{xx}\*\alpha_{E} \\
\xi_{xy}^{(1)} &= \xi_{yx}^{(1)} \approx -\gamma_{xy}\*\bigg[1-\left( \frac{9}{4}\*\gamma_{v} - 1 \right) \*\alpha_{E}^{2} \bigg]\\
\xi_{yy}^{(1)} &\approx -\gamma_{yy}\*\alpha_{E} \\
\beta_{zx}^{(2)} &= \beta_{xz}^{(2)}\approx -\frac{i\*\gamma_{xy}}{\epsilon_\perp\sqrt{1+v}}\*(1-3\*\gamma_{v})\*\alpha_{E} \\
\beta_{zy}^{(2)} &= \beta_{yz}^{(2)} \approx -i\*\gamma_{v}^{2}\*v\*\big[ 1+(1-4\*\gamma_{v})\*\alpha_{E}^{2}\big] \\
\eta_{zz}^{(2)} &\approx \frac{1}{\epsilon_{\bot}}\*\frac{v\*\gamma_{v}^{2}}{\sqrt{1+v}}\*\big[ 1+(1-4\*\gamma_{v})\*\alpha_{E}^{2}\big] \\
\xi_{xx}^{(2)} &\approx \gamma_{xx}\*\big[ 1+(1-4\*\gamma_{v})\*\alpha_{E}^{2}\big] \\
\xi_{xy}^{(2)} &= \xi_{yx}^{(2)} \approx \gamma_{xy} \left( 1- 3 \gamma_v\right)\alpha_{E}\\
\xi_{yy}^{(2)} &\approx \gamma_{yy}\*\big[ 1+(1-4\*\gamma_{v})\*\alpha_{E}^{2}\big].
\end{align}
\label{oldparam3}
\end{subequations}

\end{document}